\documentclass[floatfix,prl,aps,showpacs,superscriptaddress]{revtex4}
\usepackage{epsfig,graphicx,tabularx}
\usepackage{subfigure}
\usepackage{amsmath,amsfonts,amssymb}
\usepackage{color,braket}

\newcommand{\be}{\begin{equation}}
\newcommand{\ee}{\end{equation}}

\begin{document}

\title{Localized-Interaction-Induced Quantum Reflection and Filtering \\
of Bosonic Matter in a One-Dimensional Lattice Guide}
\author{L. Barbiero}
\affiliation{Dipartimento di Fisica e Astronomia "Galileo Galilei", Universit\`a di
Padova, 35131 Padova, Italy}
\affiliation{CNR-IOM DEMOCRITOS Simulation Center and SISSA, Via Bonomea 265 I-34136
Trieste, Italy}
\author{B. A. Malomed}
\affiliation{Department of Physical Electronics, School of Electrical Engineering,
Faculty of Engineering, Tel Aviv University, Tel Aviv 69978, Israel}
\author{L. Salasnich}
\affiliation{Dipartimento di Fisica e Astronomia "Galileo Galilei" and CNISM,
Universit\`a di Padova, 35131 Padova, Italy\\
INO-CNR, Research Unit of Sesto Fiorentino, Via Nello Carrara, 1 - 50019
Sesto Fiorentino, Italy}

\begin{abstract}
We study the dynamics of quantum bosonic waves in a one-dimensional tilted
optical lattice. An effective spatially localized nonlinear two-body
potential barrier is set at the center of the lattice. This version of the
Bose-Hubbard model can be realized in atomic Bose-Einstein condensates, with
the help of localized optical Feshbach resonance, controlled by a focused
laser beam, and in quantum optics, using an arrayed waveguide with
selectively doped guiding cores. Our numerical analysis demonstrates that
the central barrier induces anomalous quantum reflection of incident wave
packets, which acts solely on bosonic components with multiple onsite
occupancies, while single-occupancy components pass the barrier, allowing
one to distill them in the interaction zone. As a consequence, in this
region one finds a hard-core-like state, in which the multiple occupancy is
forbidden. Our results demonstrate that this regime can be attained
dynamically, using relatively weak interactions, irrespective of their sign.
Physical parameters necessary for the experimental implementation of the
setting in ultracold atomic gases are estimated.
\end{abstract}

\maketitle

\section{Introduction}

Isolated quantum systems in out-of-equilibrium configurations have attracted
a great deal of interest due to the possibility of observing new quantum
effects \cite{Polkovnikov2011}. An ideal platform to build such systems is
offered by ultracold bosons in reduced dimensionality \cite
{Luca,Delgado,Cazalilla2011}, where all parameters of the system can be
controlled with a high level of accuracy and flexibility \cite{Bloch2008}.
In this context, the band structure generated by optical lattices (OLs) \cite
{jaksch} and the absence of dissipation have allowed the experimental
observation of peculiar out-of-equilibrium effects \cite
{Winkler2006,Strohmaier2010,Mark-Haller2012} predicted several years ago
\cite{hubbard}. Atomic motion induced by tilted OL potentials has been
widely explored too, revealing remarkable quantum features \cite
{simon,meinert}. Furthermore, the study of the dynamics of bosonic waves in
a continuous geometry opens the way to a novel applications in nonlinear
optics \cite{Kivshar,Kartashov} and plasmas \cite{Stasiewicz}. Scattering of
bosonic solitary matter waves on narrow repulsive \cite{helm, martin,
cuevas, polo, helm2, helm3} and attractive \cite{ernst,lee} potential
barriers or wells has been extensively studied in a theoretical form too,
suggesting experimental observations of the effect of the quantum reflection
\cite{marchant,marchant1}. In early work \cite{Azbel} and more recently \cite
{Kartashov,Nir,students}, configurations where effective \emph{nonlinear}
potential barriers or wells are induced by spatially localized two-body
interaction have been proposed as a possible mechanism to observe other
various forms of the anomalous reflection and splitting \cite{HS}.

In this work we combine the above-mentioned ingredients to study the
scattering of wave packets, composed of non-interacting bosons in a tilted
OL, on a localized interaction zone, by means of systematic simulations
based on the time-dependent density-matrix-renormalization-group (t-DMRG)
method. Exotic effects, such as selective quantum reflection, distillation
and filtering, are revealed as a result of the scattering. In particular, we
demonstrate that, even for a relatively small interaction strength, the
nonlinear barrier acts as quantum filter, which almost completely reflects
bosonic components with multiple onsite occupancies, while the components
carrying the single occupancy (SO) are able to pass the barrier. In this
way, a region where multiple occupancies (MO) are forbidden is found. We
demonstrate that such a state can be \textit{distilled} from the incident
wave packet, using both repulsive and (rather unexpectedly) attractive
localized interactions. Furthermore, our analysis reveals that the
distillation effect, induced by the lattice's band structure, features its
most pronounced form, i.e., the total MO reflection, at relatively small
interaction strengths, and it is not essentially affected by variation of
the potential tilt which drives the incident wave packets.

\section{The model}

\begin{figure}[tbp]
\epsfig{file=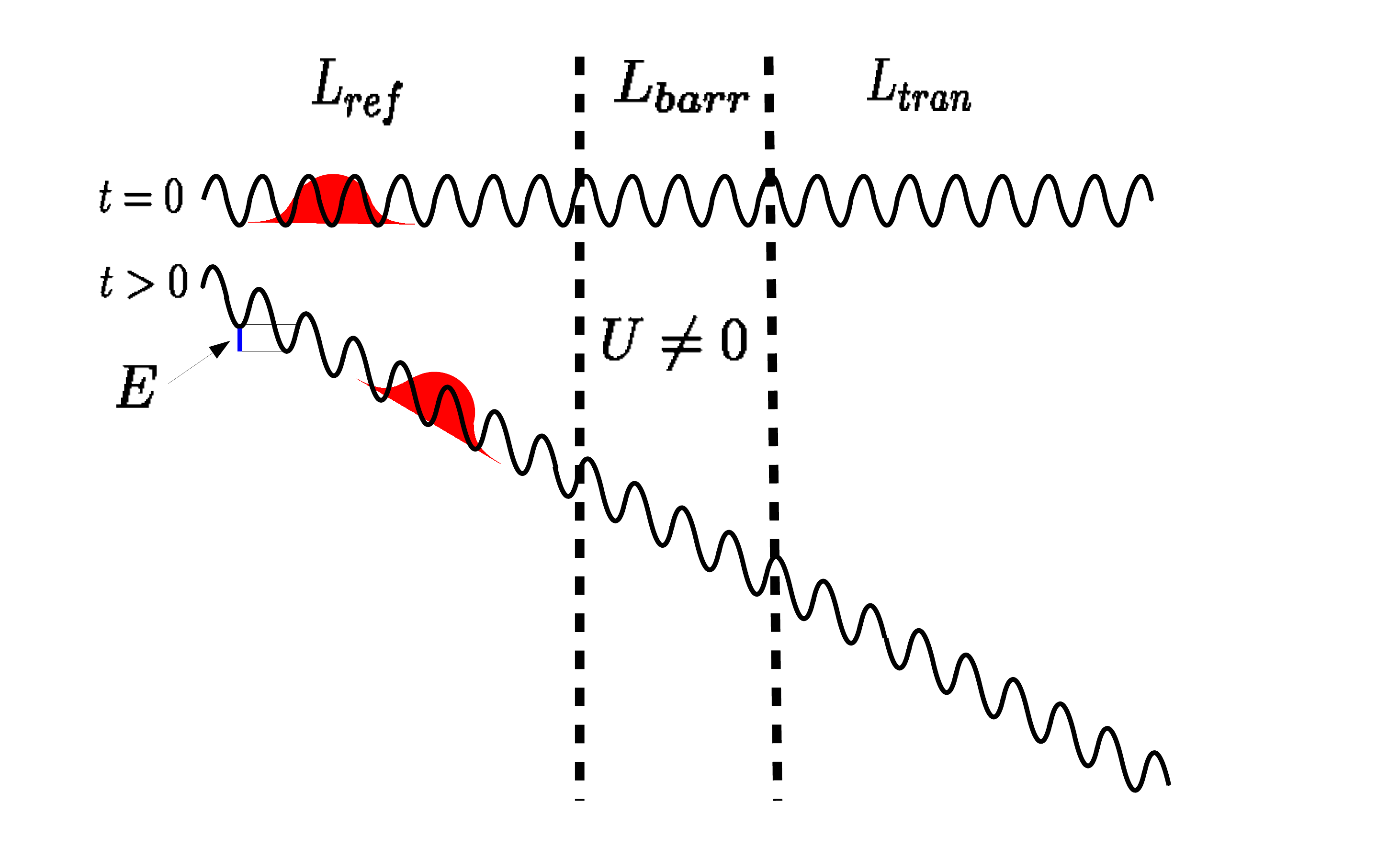,width=12cm,clip=true}
\caption{The setting under the consideration. At $t=0$, a bosonic wave
packet is placed at the edge of the left part of the lattice, of size $L_{
\mathrm{ref}}$ (in terms of the number of the OL sites), into which a part
of the wave packet is \textit{reflected} after the collision with the
interaction zone ($U\neq 0$). At $t>0$, the evolution of the input is
governed by Hamiltonian (\protect\ref{eq1}), with tilt $E$ driving the
particles towards the central interaction zone (nonlinear \textit{barrier})
of size $L_{\mathrm{barr}}$. Adjacent to it on the right-hand side, is a
part composed of $L_{\mathrm{tran}}$ OL sites, into which \textit{transmitted
} particles will move.}
\label{fig1}
\end{figure}

We study the evolution of an initially localized bosonic wave packet moving
in a one-dimensional (1D) tilted OL, with onsite interaction acting in a
finite region (``barrier") of size $L_{\mathrm{barr}}$
(measured in terms of the OL sites), where a part of incident waves may be
trapped in the case of the attractive interaction, see Fig. \ref{fig1}. At $
t=0$, we place a Gaussian wave packet near the left edge of the whole
lattice. Experimentally, the initial packet may be created by a very tight
harmonic-oscillator trap, initially applied at the same spot, which is
subsequently lifted. Potential tilt $E$ can be produced and tuned by
applying dc magnetic field along the vertical direction, with a gradient
along the OL, its effect being to induce the accelerated motion of atoms
towards the center, where a zone representing the nonlinear scatterer \cite
{Azbel,Nir,HS} is composed of a finite number of sites carrying onsite
interaction strength ($U\neq 0$).

Spatially nonuniform interactions in ultracold atomic gases were recently
realized in experiments \cite{non-uniform,non-uniform2,non-uniform3}, with
the help of the Feshbach resonance controlled by inhomogeneous external
fields. These results motivate the consideration of various settings based
on effective nonlinear potentials \cite
{Azbel,Thawatchai,Barcelona,Nir,HS,Barcelona0,boris}, which includes the
prediction of one-dimensional quantum solitons in the Bose-Hubbard (BH)
model with the strength of the onsite repulsive interaction ($U_{i}>0$)
growing with the distance from the center, $|i|,$ at any rate faster than $
|i|$ \cite{io}. On the other hand, it was shown in detail in the context of
another physical setting in Ref. \cite{HS} that a confined interaction zone,
extending over the width corresponding to a few OL sites, can be induced by
means of the optical Feshbach resonance controlled by a laser beam shone
onto the lattice in the perpendicular direction.

Here we consider the evolution of the atomic condensate governed by the
following Hamiltonian of the Bose-Hubbard (BH) type:
\begin{equation}
H=-J\sum_{i=1}^{L}(b_{i}^{\dagger }b_{i+1}+b_{i+1}^{\dagger }b_{i})+\frac{U}{
2}\sum_{i=L_{\mathrm{ref}}+1}^{L_{\mathrm{ref}}+L_{\mathrm{barr}
}}n_{i}(n_{i}-1)-E\sum_{i=1}^{L}i\,n_{i},  \label{eq1}
\end{equation}
where $b_{i}~(b_{i}^{\dagger })$ is the bosonic annihilation (creation)
operator for an atom at the $i$-th site in the lattice of total length $L$,
and $n_{i}$ is the atomic population at the site. The hopping of atoms
between nearest lattice sites is controlled, as usual, by the respective
probability $J$, which sets scales for energy and time in the present system
(i.e., $J=1$ is set below), $E$ is the potential tilt, and $U$ the strength
of the two-body interaction at those sites where it is present, i.e., $L_{
\mathrm{ref}}+1\leq i\leq L_{\mathrm{ref}}+L_{\mathrm{barr}}$. Since we
apply the interaction in a small part of the lattice, it is relevant to
distinguish three different regions: the left region with $L_{\mathrm{ref}}$
sites, into which the incident atoms are reflected, the central region of
the nonlinear barrier with $L_{\mathrm{barr}}$ sites, and the right region
with $L_{\mathrm{tran}}$ sites, into which the atoms may be transmitted.
Thus, the total number of sites in the lattice, which as a whole is embedded
into a potential box, is $L=L_{\mathrm{ref}}+L_{\mathrm{barr}}+L_{\mathrm{
tran}}$, as shown in Fig \ref{fig1}.

To estimate a possibility of the experimental implementation of the proposed
setting in ultracold gases, it is relevant to refer to recent experimental
work \cite{Innsbruck}, which used cesium atoms in the hyperfine ground
state, $\left\vert F=3,m_{F}=3\right\rangle $, for realizing regular and
chaotic regimes of the superfluid flow in tilted OLs, created by laser beams
with wavelength $\lambda =1.0645$ $\mathrm{\mu }$m, with the corresponding
depth equal to $7$ recoil energies ($E_{\mathrm{recoil}}=1.325$ kHz). This
value of the depth translates into the atom's hopping rate $J=52.3$ Hz.
Further, the scattering length $a_{s}=21.4~a_{0}$ corresponds to the onsite
interaction strength $U=102$ Hz, which, by means of the Feshbach resonance,
could be increased up to $U=533$ Hz. Thus, the setting made it possible to
easily realize values of the main control parameter, $U/J$, ranging between $
2$ and $5$. Values of this parameter which are essential to the results
reported below are virtually the same, $2<U/J<6$. The potential ramp was
created in Ref. \cite{Innsbruck} using a combination of gravity and
magnetic-field gradient, with values up to $\nabla B=31.1$ G/cm. This method
makes it possible to readily adjust values of $E/J$ to values relevant to
the present analysis, which are $0.1<E/J<0.5$, see below.

In addition to atomic Bose-Einstein condensates (BECs), the same BH system
may be implemented as a quantum-optics model of an array of evanescently
coupled parallel waveguides \cite{array} (possibly, photonic nanowires \cite
{nanowires}). In that case, the localized interaction zone can be created by
means of selectively doping the respective guiding cores by a material which
resonantly enhances the Kerr nonlinearity \cite{Kip}, while the potential
ramp can be used by tapering individual cores. In the optics model, the
evolution variable, $t$, is replaced by the propagation distance, $z$.
Typically, the hopping rate corresponds to the inter-core coupling length $
J^{-1}\lesssim 1$ cm, which makes it necessary to have the nonlinearity
length as short as $U^{-1}\sim 2$ mm. This value is challenging, but the us
of the resonantly enhanced nonlinearity may make it possible.

\section{Numerical Results}

We report numerical results obtained by means of the t-DMRG technique \cite
{Feiguin2005} using 350 DMRG states in the time-evolution calculations and
time step $\Delta t=0.01$ (it was checked that taking smaller $\Delta t$
does not affect the results). We simulated the system with $L=20$ lattice
sites and a variable number $N$ of bosons, fixing the corresponding sizes in
Eq. (\ref{eq1}) as $L_{\mathrm{ref}}=L_{\mathrm{tran}}=8$ and $L_{\mathrm{
barr}}=4$. Although the total size used here, $L=20$, is relatively small,
it is comparable with that in experimentally realized systems \cite{magnons}
. It can be checked that the increase of $L$ and, accordingly, of $L_{
\mathrm{ref}}$, $L_{\mathrm{barr}}$, $L_{\mathrm{tran}}$ affects only
characteristic time scales of the dynamical results reported below, but does
not essentially alter outcomes of the scattering.

\subsection{The Flat Repulsive Barrier}

\begin{figure}[tbp]
\epsfig{file=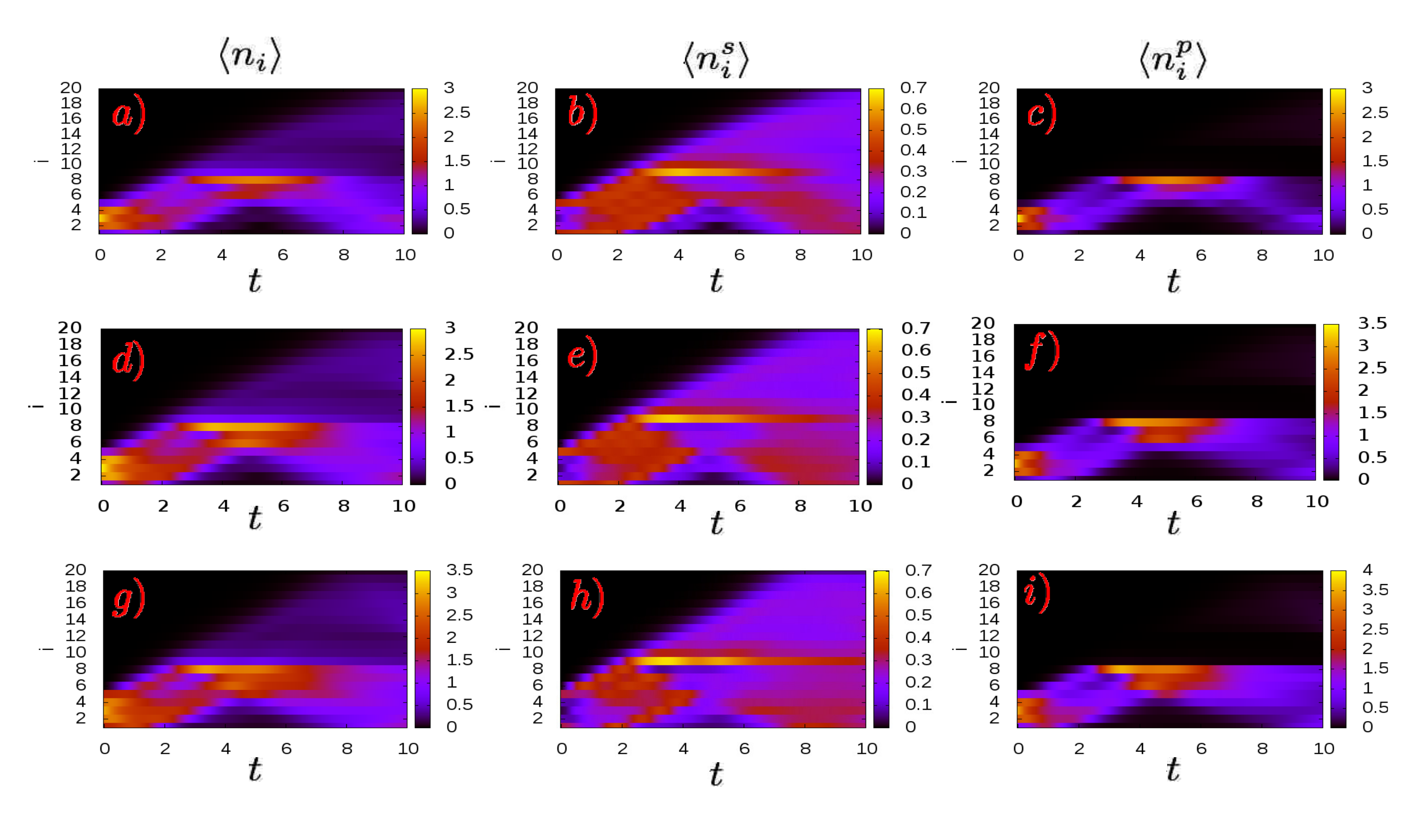,width=16cm,clip=true}
\caption{The first, second and third columns show the evolution of $
\left\langle n_{i}\right\rangle $, $\left\langle n_{i}^{s}\right\rangle $,
and $\left\langle n_{i}^{p}\right\rangle $, respectively. The single- and
multiple-occupancy expectation values, $\left\langle n_{i}^{s}\right\rangle $
and $\left\langle n_{i}^{p}\right\rangle $, are produced by the operators
defined in Eqs. (\protect\ref{s}) and (\protect\ref{p}), respectively. They
are evaluated in the system with $N=8,10,$ and $12$ bosons, in the first,
second and third rows, respectively. All the results refer to a lattice with
$L=20$ sites, with $N$ bosons placed at $t=0$ on 5 lattice sites. The other
parameters are $L_{\mathrm{ref}}=L_{\mathrm{tran}}=8$, $L_{\mathrm{barr}}=4$
, $U=6.0$, and $E=0.3$. }
\label{fig2}
\end{figure}

Usually, reflection and transmission of wave packets is revealed by tracking
expectation values $\left\langle n_{i}\right\rangle $ of the density profile
with respect to the evolving many-body quantum state, $\left\vert \psi
(t)\right\rangle $. Note that $\left\langle n_{i}\right\rangle $ can be
precisely measured in the experiment, by means of the recently developed
in-situ imaging technique \cite{greiner,in-situ}. The Gaussian shape of $
\left\langle n_{i}\right\rangle $ at $t=0$ is localized on five lattice
sites populated by $N=8,10,12$ bosons, respectively, in the first, second
and third row of Fig. \ref{fig2}. Once at $t>0$ the bosons are free to move
toward the central part of the system, the initial Gaussian density profile
is deformed \cite{noteGauss}, and its actual shape depends on $N$, as is
evident in the first column of Fig \ref{fig2}.

Figures \ref{fig2}$a$), $d$) and $g$) show typical quantum-reflection
effects for $E=0.3$ and $U=6.0$. Indeed, it is seen that, as the bosons
approach the interaction zone with $U\neq 0$ \cite{note}, a large fraction
of them bounce back, with only a small part being able to pass $L_{\mathrm{
barr}}$. At the first glance, this behavior is similar to the one induced by
the usual linear potential barrier, see, e.g., Ref. \cite{marchant1}.
However, a crucial difference is that the \emph{nonlinear}
(interaction-induced) barrier in our setting acts only on two- and many-body
states. Therefore, it is necessary to distinguish between the SO and MO
scattering behaviors. To this end, we define two operators acting on the
many-body quantum state, $\left\vert \psi (t)\right\rangle $. One operator
counts SO sites, with occupancy $\left\langle n_{i}\right\rangle \leq 1$:
\begin{equation}
n_{i}^{s}\left\vert \psi (t)\right\rangle =\alpha \left\vert \psi
(t)\right\rangle \quad \mathrm{with}\quad \alpha =\left\langle
n_{i}\right\rangle \quad \mathrm{if}\quad \left\langle n_{i}\right\rangle
\leq 1,~\mathrm{and}\quad \alpha =0\quad \mathrm{if}\quad \left\langle
n_{i}\right\rangle >1.  \label{s}
\end{equation}%
The operator counting MO sites is $n_{i}^{p}=n_{i}(n_{i}-1)/2$, which acts
according to
\begin{equation}
n_{i}^{p}\left\vert \psi (t)\right\rangle =\beta \left\vert \psi
(t)\right\rangle \quad \mathrm{with}\quad \beta =0\quad \mathrm{if}\quad
\left\langle n_{i}\right\rangle \leq 1,~\mathrm{and}\quad \beta
=\left\langle n_{i}(n_{i}-1)\right\rangle /2\quad \mathrm{if}\quad
\left\langle n_{i}\right\rangle >1.  \label{p}
\end{equation}%
Thus we can separately take into account sites where the interaction, if
present, is effective, i.e. $\left\langle n_{i}^{p}\right\rangle \neq 0$,
and where it is not, i.e., $\left\langle n_{i}^{s}\right\rangle \neq 0$. In
the second and third columns of Fig. \ref{fig2}, respectively, we show the
evolution of the expectation values of operators $n_{i}^{s}$ and $n_{i}^{p}$
. While it is evident from Figs. \ref{fig2}$b$), $e$), and $h$) that the SO,
represented by $\left\langle n_{i}^{s}\right\rangle $, freely passes the
nonlinear barrier, Figs. \ref{fig2}$c$), $f$), and $i$) make it clear that
the MO, represented by $\left\langle n_{i}^{p}\right\rangle $, bounces back
from it. More precisely, we notice that, after an initial decrease due to
the propagation in the non-interaction regime, $\left\langle
n_{i}^{p}\right\rangle $ consistently grows at the right edge of $L_{\mathrm{
ref}}$ at intermediate times. The accumulation of the MO is followed by its
nearly complete rebound. On the other hand, the SO features the behavior
reverse to that of $\left\langle n_{i}^{p}\right\rangle $ in the $L_{\mathrm{
ref}}$ section of the lattice. In particular, dissociation (formation) of
the MO is coupled to the increase (decrease) of $\left\langle
n_{i}^{s}\right\rangle $.

\begin{figure}[tbp]
\epsfig{file=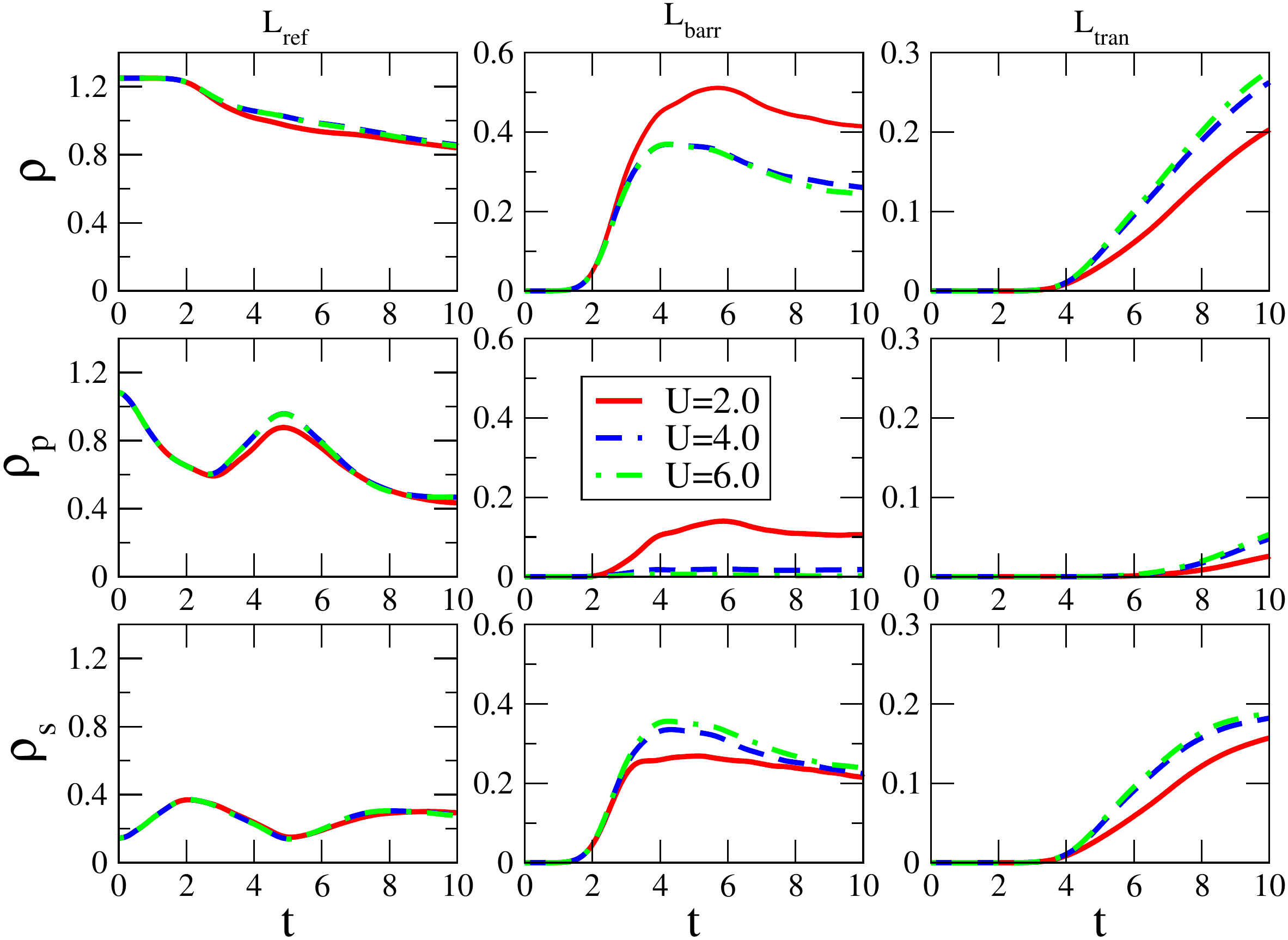,width=10cm,clip=true}
\caption{The three rows display, respectively, the evolution of average
densities of the total number of bosons, SO (single occupancy), and MO
(multiple occupancy), in the three sections of the lattice, which are
defined as per Eq. (\protect\ref{rho}). The three columns refer to sections $
L_{\mathrm{ref}}$, $L_{\mathrm{barr}}$ and $L_{\mathrm{tran}}$, as indicated
in the top line. The data were collected for $N=10$ bosons and $L=20$ sites,
with $N$ bosons placed, at $t=0$, on 5 lattice sites. The other parameters
are $L_{\mathrm{ref}}=L_{\mathrm{tran}}=8$, $L_{\mathrm{barr}}=4$, $E=0.3$,
and different values of $U$, as indicated in the figure.}
\label{fig5}
\end{figure}

The crucial point is the behavior of the bosons inside the central
interaction zone, where, on the contrary to the MO, the SO can evidently
reside. This fact is a drastic difference with respect to the usual
settings, with a linear-potential barrier acting at the single-particle
level. Thus, as already stated, the quantum transmission observed in Figs. 
\ref{fig2}$a$), $d$), and $g$) is totally accounted for by the SO motion. In
other words, the interaction zone acts as a \textit{quantum filter}, which
sends all the occupancies with $\left\langle n_{i}\right\rangle >1$ back,
and lets those with $\left\langle n_{i}\right\rangle <1$ pass. In this way,
the interaction zone, cleared of the MO, displays an effective hard-core
on-site repulsion, with bosonic particles emulating fermions. A quantum gas
where the pair- and multiple-occupations are forbidden due to the
interaction is usually associated to the appearance of the Tonks-Girardeau
(TG) regime. The latter was originally predicted in a configuration
preserving the Galilean invariance \cite{tonks,bijl,girardeau}, but it has
later been demonstrated both theoretically \cite{cazalilla2003} and
experimentally \cite{paredes} that the presence of a lattice preserves the
main features of the TG gas. Noticeably, in Fig. \ref{fig2} the hard-core
constraint is generated for all considered values of the boson number, $N$.
The latter fact signals that the interaction strength, $U$, is responsible
for the filtering effect. To check the efficiency of the filter, in Fig. \ref
{fig5} we plot densities which are, respectively, the observation values of $
n_{i}$, $n_{i}^{p}$ and $n_{i}^{s}$, averaged over three different parts of
the lattice, $L_{\mathrm{r,t,t}}\equiv \left\{ L_{\mathrm{ref}},L_{\mathrm{
barr}},L_{\mathrm{tran}}\right\} $, for different values of the interaction
strength, $U$:%
\begin{equation}
\rho =L_{\mathrm{r,b,t}}^{-1}\sum_{L_{\mathrm{r,b,t}}}\left\langle
n_{i}\right\rangle ,\quad \rho _{p}=L_{\mathrm{r,b,t}}^{-1}\sum_{L_{\mathrm{
r,b,t}}}\left\langle n_{i}^{p}\right\rangle ,\quad \rho _{s}=L_{\mathrm{r,b,t
}}^{-1}\sum_{L_{\mathrm{r,b,t}}}\left\langle n_{i}^{s}\right\rangle
\label{rho}
\end{equation}
It is clearly seen in Fig. \ref{fig5} that, in the course of the evolution
the value of $\rho _{p}$ is conspicuously different from zero in region $L_{
\mathrm{barr}}$ only for a relatively weak interaction strength, namely, $
U=2 $. Once a stronger interaction acts in $L_{\mathrm{barr}}$, the MO
density practically vanishes. As a result, at intermediate values of time, a
gas composed of the SO is stabilized in the distilled form inside the
interaction zone. Obviously, the number of particles approaching the barrier
does not depend on the interaction strength present in $L_{\mathrm{barr}}$,
as is evident in the first column of Fig. \ref{fig5}. In the same time, once
the interaction capable to support the quantum filtering is applied in $L_{
\mathrm{barr}}$, the value of $\rho _{s}$ becomes independent of the
interaction strength $U$. This is confirmed by the fact that in Fig. \ref
{fig5} we see that, for $U=4$ and $6$, $\rho _{p}$ is actually zero in
region $L_{\mathrm{barr}}$, hence $\rho _{s}$ has the same value for these
two interaction strengths. As seen in the third column of Fig. \ref{fig5},
this aspect has its consequences also in the behavior of bosons in region $
L_{\mathrm{tran}}$. Indeed, the filtering process allows a larger number of
bosons to enter $L_{\mathrm{tran}}$, which means that, effectively, the
dynamically induced hard-core constraint increases the speed of the
particles. In particular, our measurements yield $\rho (U=4,6)/\rho
(U=2)\approx 1,25$ at $t=10$. Interestingly this larger amount of particles
allows the formation of higher SO and MO alike, see values of $\rho _{s}$
and $\rho _{p}$ in the third column of Fig. \ref{fig5}.

\subsection{The Flat Attractive Barrier}

\begin{figure}[tbp]
\epsfig{file=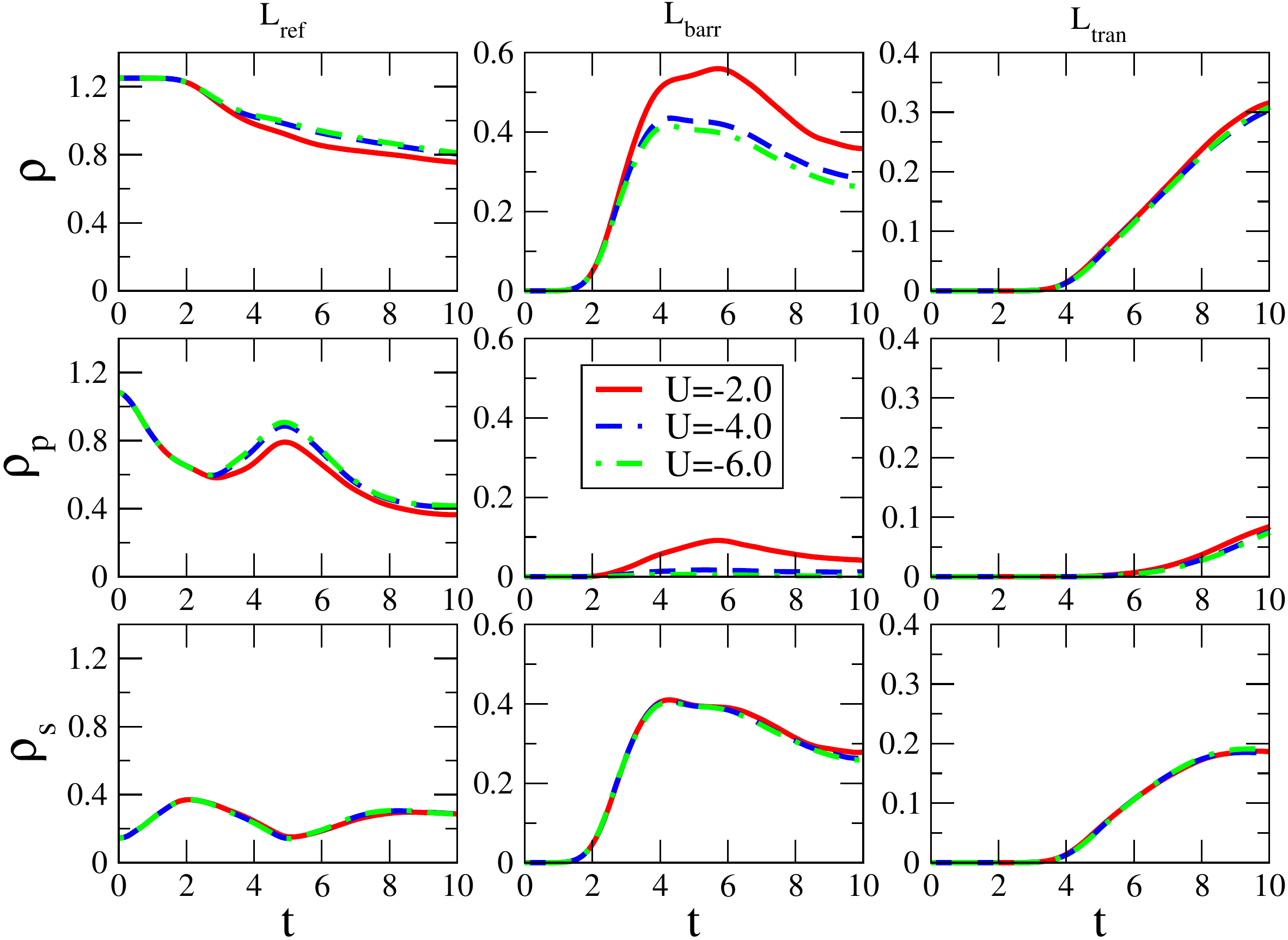,width=10cm,clip=true}
\caption{The same as in Fig. \protect\ref{fig5}, but for the system with the
attraction ($U<0$) acting in the interaction zone ($L_{\mathrm{barr}}$).}
\label{fig6}
\end{figure}

The quantum reflection of the MO might seem a rather obvious consequence of
the repulsive nature of the interaction. For this reason, it is interesting
to consider the system with attractive interactions, i.e., $U<0$, too. The
analysis of static configurations for $U<0$ and relatively large $|U|$ has
previously revealed collapsed states, see Refs. \cite{Nir,io2} and
references therein. This fact suggests that MO may not bounce back from the
interaction zone, $L_{\mathrm{trap}}$, and get partly trapped in it.
Nevertheless, Fig. \ref{fig6}, which displays the same characteristics of
the dynamical scattering as in Fig. \ref{fig5}, but for $U<0$, shows that
this \emph{does not happen} -- in fact, the self-attraction zone does not
accumulate the MO. Actually we observe that this system again stabilizes an
effectively ``distilled" quasi-TG state in this zone,
although with a higher density than in the case of $U>0$.

The approximate symmetry between the cases of $U>0$ and $U<0$, revealed by
the comparison of Figs. \ref{fig5} and \ref{fig6}, agrees with findings of
Ref. \cite{schneider}, where a similar symmetry was discovered in the
transport of fermion atoms. In the present contexts, it is related to
properties of the energy spectrum of lattice bosons, which demonstrates the
symmetry with respect to $U\leftrightarrow -U$. Moreover, the consideration
of the attractive interaction helps one to understand how the present BH
model gives rise to the quantum filtering, distillation, and rebound
effects. First, it is obvious that, in either case, the system conserves the
total energy (along with the total number of bosons). Further, the band
structure produced by the OL imposes a limitation on possible values of the
kinetic energy. In fact, the formation of MO in the interaction zone would
induce energy variation that cannot be supported by the system in which any
gain/loss in the potential energy must be converted into the opposite change
of the kinetic energy. A precise many-body quantification of this effect is
a very hard problem, due to the non-integrability of eq. \ref{eq1}.
Nevertheless, arguments regarding the two- \cite
{Winkler2006,Strohmaier2010,Mark-Haller2012} and three-body \cite{johnson}
bound states may be sufficient to explain many significant dynamical quantum
effects in 1D lattice systems \cite
{Winkler2006,Strohmaier2010,Mark-Haller2012,io3}. In our case, the
derivation of the two- and three-body energy spectrum is substantially
complicated by the presence of the tilted potential. Nevertheless, the same
energy arguments make it possible to explain the above-mentioned effect.

Actually, the rebound of the MO bosonic component from the self-attraction
region, observed in Fig. \ref{fig6}, is alike to the commonly known effect
of the partial reflection of an incident wave from a quantum-mechanical
potential well \cite{LL}, and also to the possibility of the rebound of a
moving soliton from a potential well in the nonlinear model \cite{Brand}.

\subsection{A Linearly Shaped Repulsive Barrier}

All the results presented above refer to a configuration where the bosons
are subject to spatially uniform interactions in region $L_{\mathrm{barr}}$.
A essential issue is whether a barrier with spatially inhomogeneous
interactions gives rise to similar filtering effects.
\begin{figure}[tbp]
\epsfig{file=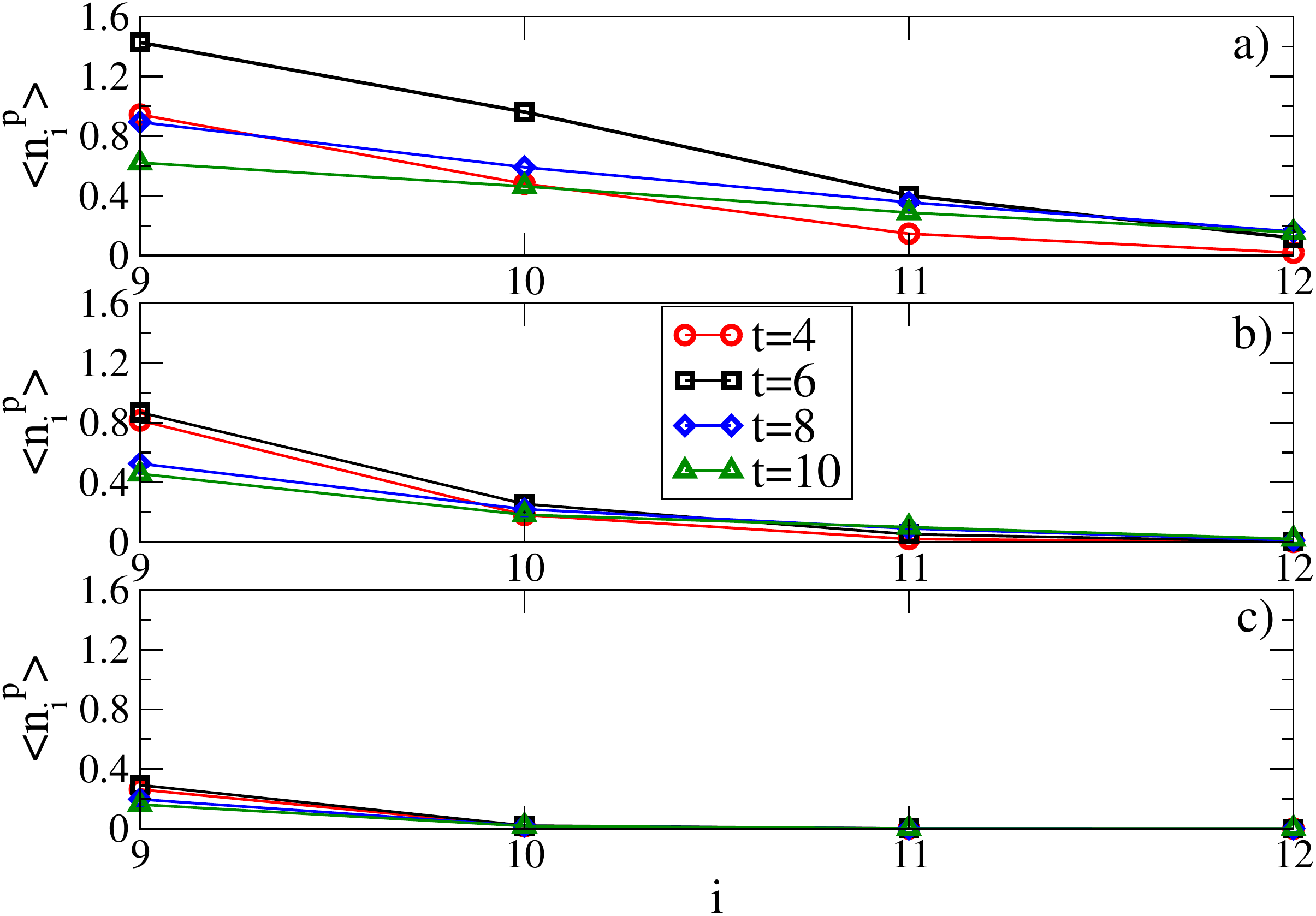,width=10cm,clip=true}
\caption{The expectation value of $\langle n_{i}^{p}\rangle $ in region $L_{
\mathrm{barr}}$, namely, at sites $i=9,10,11,12$, in the system with $N=10$
bosons initially placed on $5$ lattice sites, and $E=0.3$. The local
interaction strength, $U(i)=U_{\min }\cdot (i-8)$, grows linearly with the
distance, with slope $U\min =0.5,1.0,$ and $2.0$ in $a$), $b$) and $c$),
respectively.}
\label{figslow}
\end{figure}

In Fig. \ref{figslow} we show the behavior of $\langle n_{i}^{p}\rangle $,
evaluated at different times in region $L_{\mathrm{barr}}$ in the system
with the interaction strength growing linearly with the distance. More
precisely, we study a configuration where the bosons are subject to the
site-dependent interaction $U(i)$, with minimum value $U_{\min }$ at site $
i=9$, and maximum strength $U_{\max }$ at $i=12$, i.e. $U(i)=U_{\min }(i-8)$
. As might be expected, it is observed in Fig. \ref{figslow} $a$),
corresponding to $U_{\min }=0.5$, that such a weak potential is not able to
support any filtering. Noticeably, at site $i=12$, where the strength is $
U(i=12)=2$, we find that $\langle n_{i}^{p}\rangle $ has the same value of $
\rho _{p}$ as evaluated in region $L_{\mathrm{barr}}$ in Fig. \ref{fig5} for
$U=2$ \cite{note3}. A similar feature is shown in Fig. \ref{figslow} $b$),
where $U_{\min }=1$. Here we note that, at site $i=10$, where $U(10)=2$, we
again find the same value of $\rho _{p}$, averaged over $L_{\mathrm{barr}}$,
as in Fig. \ref{fig5} for $U=2$. Moreover, it is relevant to point out that
the only site where MO is actually forbidden is the point where $U(i=12)=4$,
again in agreement with Fig. \ref{fig5} for $U=4$. Finally, the same
correspondence with Fig. \ref{fig5} is observed in in Fig. \ref{figslow} $c$
), where $U_{\min }=2$. Here, the strong interaction produces filtering
effects at all sites but $i=9$, where the local interaction strength is not
strong enough, $U(9)=2$. Notice that the number of particles and
single-/multiple-occupations at $i>12$ is exactly the same as in the case of
the flat barrier. In particular, the state at $i>12$ for $U_{\min }=0.5$ is
exactly the same as that observed in column 3 of Fig. 3 \ (the red curve).

Finally, we make conclusions for the present case. First , we conclude, as
expected, that the only ingredient generating the filtering is the strength
of the interaction, but not its spatial distribution. Moreover, the
comparison of Figs. \ref{figslow} and \ref{fig5} makes it clear that the
size of the interaction zone does not play any role in the generation of the
effective hard-core constraint on the MO. More precisely, the larger is the
number of sites with sufficiently strong interaction, the broader is the
region where the effective hard-core repulsion is present.

\subsection{Different Initial Configurations}

\begin{figure}[tbp]
\epsfig{file=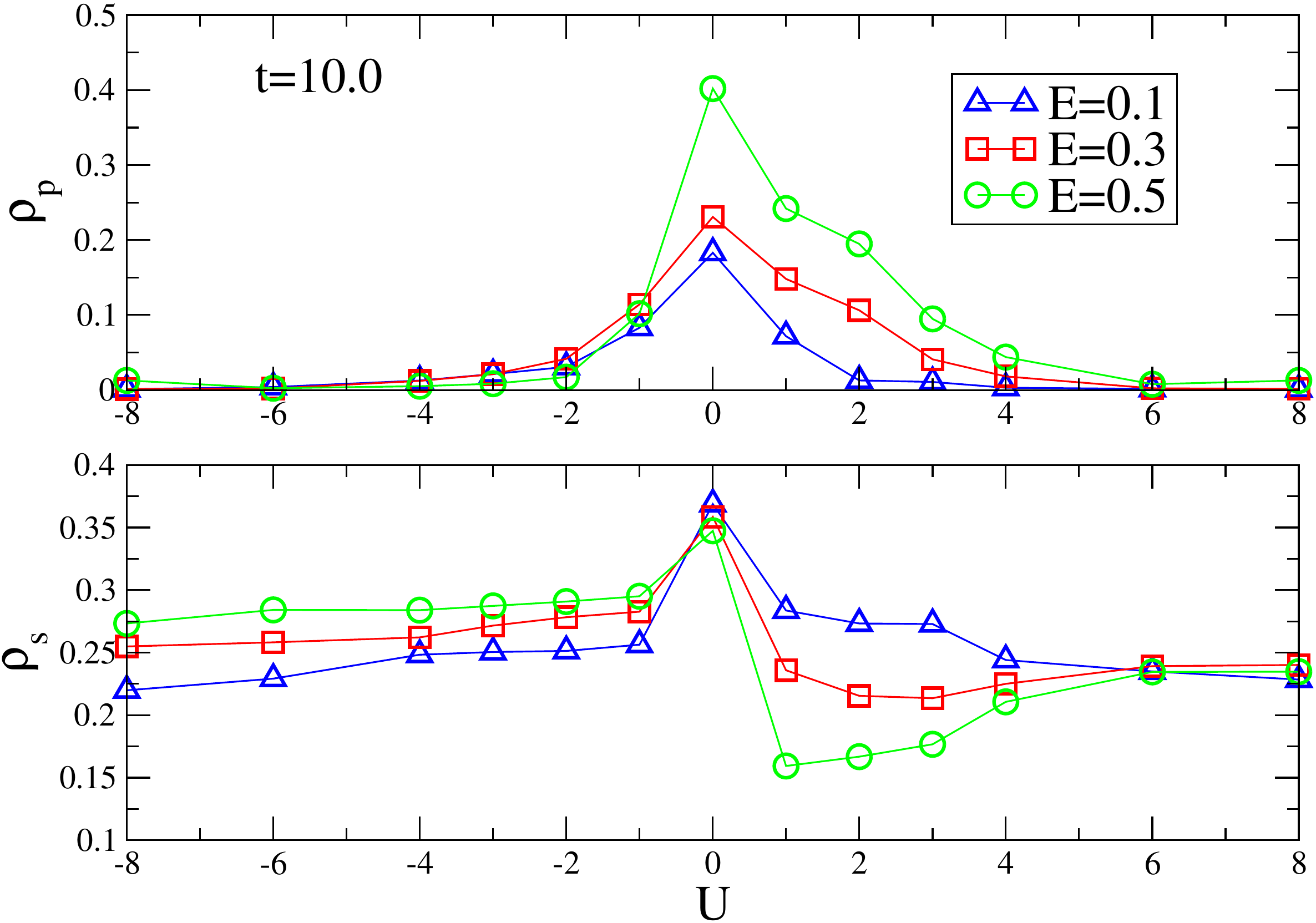,width=10cm,clip=true}
\caption{Average MO and SO densities, $\protect\rho _{p}$ and $\protect\rho
_{s}$, in the interaction zone, defined as per Eq. (\protect\ref{rho}) (with
$L_{\mathrm{r,t,t}}=L_{\mathrm{barr}}$), taken at $t=10$, as functions of
the interaction strength $U$, at different fixed values of the potential
tilt, $E$. Initially, $N=10$ bosons are placed on 5 lattice sites.}
\label{fig8}
\end{figure}

As pointed out above, the energy considerations determine the filtering and
reflection effects outlined above. Actually, the energy of the initial state
in eq. (\ref{eq1}) depends on several parameters, such as the potential
tilt, $E$, the spatial extension of the wave packet at $t=0$, and the number
of bosons, $N$, which is the crucial quantity controlling the effects
described above. For this reason, it is relevant to explore how different
initial configurations affect our findings. As clearly seen in Fig. \ref
{fig1}, a small variation of $N$ does not bring any conspicuous variation in
the filtering properties. A different role is played by $E$. The respective
results are displayed in Fig. \ref{fig8}, where we plot the average MO and
SO densities, $\rho _{p}$ and $\rho _{s}$, in the interaction zone for
different values of $E$ at a fixed evolution time, $t=10$. The figure
clearly shows that the MO density in the interaction zone is conspicuously
affected by $E$ only at sufficiently small values of the interaction
strength, $|U|$ \cite{note2}, while $\rho _{p}$ practically vanishes at
larger values of $|U|$.

\begin{figure}[tbp]
\epsfig{file=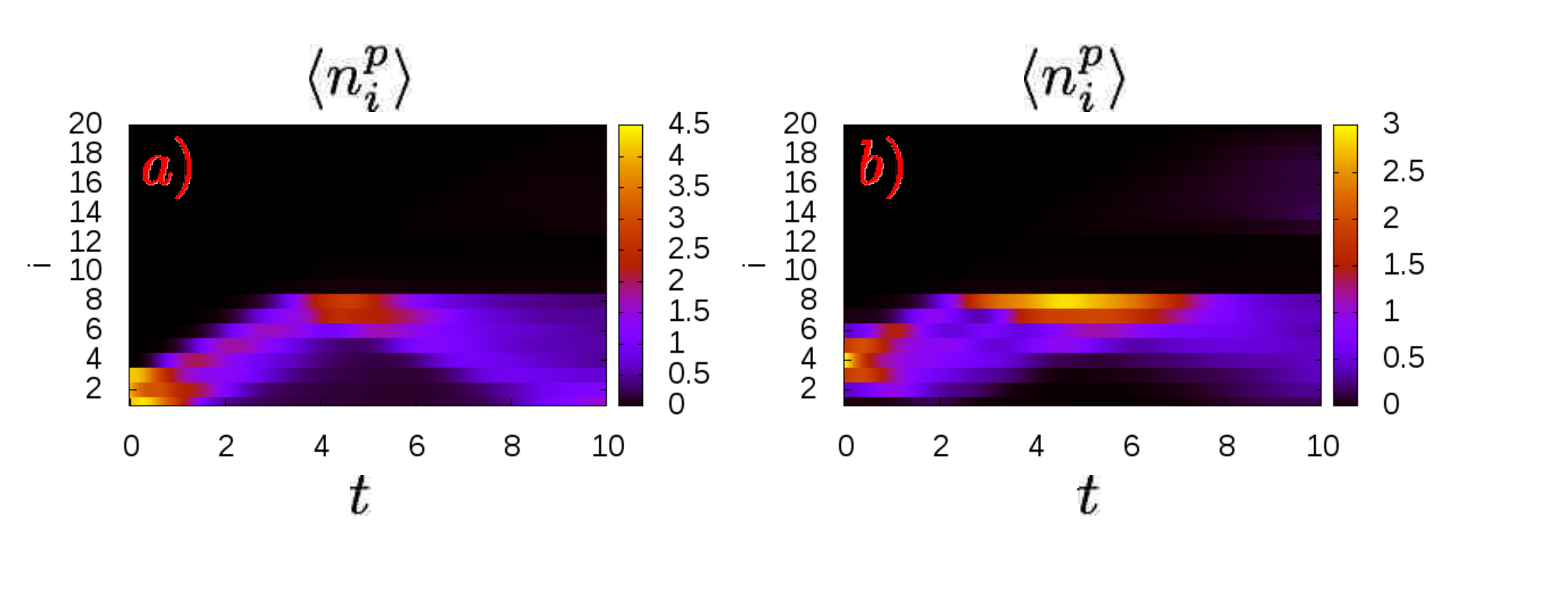,width=17cm,clip=true}
\caption{Evolution of $\langle n_{i}^{p}\rangle $ in the system with $N=10$,
$E=0.3$, and $U=6$ in the central part of the lattice. $a$) and $b$) refer
to a configuration where the initial Gaussian profile is localized over 3 or
7 sites, respectively.}
\label{fig9}
\end{figure}

On the other hand, the average SO density in the interaction zone, $\rho
_{s} $, shows a weak dependence on $E$ at almost all values of $U$, as shown
in the lower panel of Fig. \ref{fig8}, suggesting that $E$ can be used to
adjust the density of the TG state ``distilled" in the
interaction zone. We thus conclude that the effects of the quantum filtering
and MO reflection persist in the present version of the BH system at small
and intermediate values of $E$. On the contrary, the situation becomes
trivial at large $E$, when the potential ramp becomes a dominant factor
determining the dynamics of the wave packets. Finally, in Fig. \ref{fig9} we
display the evolution of $\langle n_{i}^{p}\rangle $ for wave packets
initially localized on different numbers of sites, but with the same number
of particles, $N=10$.

Thus, we infer that the variation induced by a difference in the
localization of the initial density profile does not cause any appreciable
modification of the physical features described above. Of course, much
larger variations in the number of the initially occupied sites may alter
the critical value of $U$ which gives rise to the effective hard-core
behavior. Nevertheless, we conclude that, due to energy considerations, it
is always possible, in the one-dimensional isolated quantum system, to find
a critical strength of the interaction able to give rise to the filtering
precesses.

\section{Conclusion}

We have introduced a version of the Bose-Hubbard system composed of two
sections which do not carry onsite two-body interactions, with an
interaction zone sandwiched between them. The cases of spatially uniform
repulsive and attractive interactions, as well as inhomogeneous
interactions, were considered. This is a fully quantum counterpart of models
with nonlinear potential barriers or wells, that were recently studied in
optics and mean-field description of matter waves in atomic BEC. In those
contexts, the spatially localized interactions may be induced, respectively,
by means of selective doping, or by the Feshbach resonance controlled by an
inhomogeneous external field. Using the quasi-exact numerically implemented
t-DMRG method, we have considered the scattering problem, where the
potential tilt sends a wave packet to collide with the effective nonlinear
barrier (interaction zone). The result is that the nonlinear barrier, being
transparent to the bosonic-wave component with the onsite SO (single
occupancy), induces strong quantum reflection of the MO (multiple-occupancy)
bosonic components. These properties make it possible to realize the quantum
distillation of the SO component in the interaction zone, which is
tantamount to inducing an effective on-site hard-core repulsion. The absence
of the MO, which was experimentally demonstrated to be a characteristic
feature of the TG state \cite{kinoshita,paredes}, makes it possible to
dynamically realize a similar state in the interaction zone of the present
system. Furthermore, we have shown that, in contrast to the static
configuration, where the hard-core regime occurs for very strong repulsive
interaction (while strong attraction may generate a highly excited state in
the form of the super-TG gas \cite{super1,super2,super3,super4}), our
dynamical setting makes it possible to reach this hard-core-like regime,
using relatively weak repulsion, or even weak attraction (which is an
unexpected finding), in the interaction zone. Further investigations are
currently in progress, to better characterize this peculiar regime. It has
been demonstrated that the predicted results can be implemented using
currently available experimental settings.

\section{Acknowledgments}

We appreciate a valuable discussion with M. D. Lukin. This work was
supported by MIUR (FIRB 2012, Grant No. RBFR12NLNA-002; PRIN 2013, Grant No.
2010LLKJBX). L.B. thanks CNR-INO BEC Center in Trento for CPU time.


\begin{thebibliography}{99}
\bibitem{noteGauss} In a tilted lattice the motion of a wave packet with a
permanent shape is impossible due to the constrain imposed by the energy
conservation.
\bibitem{note}In all the cases shown the $U\neq 0$ term is located only at sites with $%
i=9,10,11,12$.
\bibitem{note3} Note that we are here referring to the value of $\langle
n_{i}^{p}\rangle $ at the $i$-th site. It actually corresponds to having the
barrier with interaction constant $U=2$ on one lattice site. For this
reason, we have $\langle n_{i}^{p}\rangle =\rho _{p}$ following the
definition of \ref{rho}.
\bibitem{note2} Actually we observe that the value of $\rho _{p}$ is more
affected by $U$ in the case of the repulsive interaction.
\end{thebibliography}
\end{document}